\documentclass{aa}  
\usepackage{graphicx}
\usepackage{xcolor}
\usepackage{txfonts}
\usepackage{hyperref}

\definecolor{seagreen}{rgb}{0.190, 0.525, 0.361}
\definecolor{cerulean}{rgb}{0.165, 0.322, 0.745}
\definecolor{goldenrod}{rgb}{0.855, 0.647, 0.125}

\begin{document}

   \title{Chaos in violent relaxation dynamics}

   \subtitle{Disentangling micro- and macro-chaos in numerical experiments of dissipationless collapse}

   \author{Simone Sartorello
          \inst{1,2}
          \and
          Pierfrancesco Di Cintio\inst{3,4,5}
          \and
          Alessandro Alberto Trani\inst{6,7}
          \and
          Mario Pasquato\inst{8,9} 
          }
   \institute{Consorzio RFX (CNR, ENEA, INFN, Università di Padova, Acciaierie Venete SpA), Corso Stati Uniti 4, 35127 Padova, Italy
   \and
Centro Ricerche Fusione, Università di Padova, Padova, Italy\\ 	
	\email{simone.sartorello@phd.unipd.it}
	\and
	Consiglio nazionale delle Ricerche, Istituto dei Sistemi Complessi Via Madonna del piano 10, I-50019 Sesto Fiorentino, Italy
	\and
	Istituto Nazionale di Fisica Nucleare - Sezione di Firenze, via G. Sansone 1, I-50019 Sesto Fiorentino, Italy
	\and
	Istituto Nazionale di Astrofisica - Osservatorio Astrofisico di Arcetri, Piazzale E.\ Fermi 5, I-50125 Firenze, Italy\\
 \email{pierfrancesco.dicintio@cnr.it}
  \and
 Niels Bohr International Academy, Niels Bohr Institute, Blegdamsvej 17, 2100 Copenhagen, Denmark
    \and
     Istituto Nazionale di Fisica Nucleare - Sezione di Trieste, I-34127, Trieste, Italy\\
     \email{aatrani@gmail.com}
     \and
     Istituto Nazionale di Astrofisica - IASF Via Alfonso Corti 12 I-20133 Milano, Italy
 \and Istituto Nazionale di Astrofisica - Osservatorio Astronomico di Padova, Vicolo dell'Osservatorio 5, I-35122 Padova, Italy\\
 }

   \date{Received 28 March 2025; Accepted 11 April 2025}

  \abstract
   {} 
   {Violent relaxation (VR) is often regarded as the mechanism leading stellar systems to collisionless meta equilibrium via rapid changes in the collective potential.}
   {We investigate the role of chaotic instabilities on single particle orbits in leading to nearly-invariant phase-space distributions, aiming at disentangling it from the chaos induced by collective oscillations in the self-consistent potential.}
   {We explore as function of the systems size (i.e. number of particles $N$) the chaoticity in terms of the largest Lyapunov exponent of test trajectories in a simplified model of gravitational cold collapse, mimicking a $N-$body calculation via a time dependent smooth potential and a noise-friction process accounting for the discreteness effects. A new numerical method to evaluate effective Lyapunov exponents for stochastic models is presented and tested.}
   {We find that the evolution of the phase-space of independent trajectories reproduces rather well what observed in self-consistent $N-$body simulations of dissipationless collapses. The chaoticity of test orbits rapidly decreases with $N$ for particles that remain weakly bounded in the model potential, while it decreases with different power laws for more bound orbits, consistently with what observed in previous self-consistent $N$-body simulations. The largest Lyapunov exponents of ensembles of orbits starting from initial conditions uniformly sampling the accessible phase-space are somewhat constant for $N\lesssim 10^9$, while decreases towards the continuum limit with a power-law trend. Moreover, our numerical results appear to confirm the trend of a specific formulation of dynamical entropy and its relation with Lyapunov time scales.}
   {}


   \maketitle
%
%-------------------------------------------------------------------

\section{Introduction}
Collisionless stellar systems are commonly thought to reach equilibrium distributions via the process of violent relaxation (VR, \citealt{Lynden-Bell:1966zjv}), associated to strong variations of the collective potential, happening for example during gravitational collapse or merger events (see \citealt{binney}). In practice, strong variations of the self-consistent gravitational potential $\Phi$ happening on a time scale of the order of a few crossing times $t_{\rm c}\approx (G\tilde{\rho})^{-1/2}$ efficiently induce a redistribution of particles' energies (\citealt{1998NYASA.848...28K}) that bring the system towards a nearly invariant equilibrium distribution, often dubbed quasi-stationary state (QSS, e.g. see \citealt{2014plis.book.....C}) in the context of long-range interactions.\\
\indent Since its original inception, VR has been widely studied with a statistical mechanics approach (e.g. see \citealt{1980Ap&SS..72..293S,1986MNRAS.219..285T,1987ApJ...316..497M,1987MNRAS.225..995K,1990PhR...188..285P,1996ApJ...457..287S,2000ApJ...531..739N,2005MNRAS.361..385A,2011A&A...526A.147K,2013IJMPB..2761011H}, and references therein). In particular, in the continuum idealized limit, the approach towards the QSS of the phase-space distribution function $f$ as governed by the collisionless Boltzmann equation (CBE), is interpreted adopting different choices of coarse-graining in phase-space (see \citealt{1998ApJ...500..120K,2005MNRAS.360..892D,2014PhR...535....1L,2022ApJ...935..135B,2022MNRAS.512.3015B,2024PhRvE.109e4118W}, see also \citealt{2024arXiv240707947B} and references therein for a complete discussion) to construct an effective dissipative process over a conservative dynamics. This stems from the fact that in the CBE-governed dynamics, the phase-space distribution $f$ evolves as in incompressible fluid. Each continuum function $\mathcal{G}(f)$ integrated over the accessible phase-space (i.e. the so-called Casimir invariants) is therefore conserved and thus any entropy in the form $S=\int f\ln f{\rm d}r{\rm d}v$ does not evolve. This prevents the use of entropy maximization arguments to describe the evolution towards the microcanonical equilibrium. Recently, \cite{2017ApJ...846..125B,2019ApJ...872...20B,2019ApJ...870..128B} conjectured that the continuum limit approach based on the CBE breaks down during processes of VR and stated that the fast evolution towards meta-equilibria is also driven by the so-called chaotic instabilities. Such instability were conjectured to significantly depend on the numbers of degrees of freedom $N$ (i.e. the number of particles of the system), even for astrophysically significant sizes of $N\approx 10^{11}$, for which the use of continuum arguments is typically accepted.\\
\indent The role discreteness effects and $N-$body chaos in driving the evolution of collisionless gravitating systems has been acknowledged since at least the work of \cite{1964ApJ...140..250M,1971JCoPh...8..449M} on the irreversibility of the $N-$body dynamics in small $N$ systems. Moreover, several authors (\citealt{GurzadianSavvidy,1992A&A...266..215V,1993ApJ...415..715G,1995PhRvE..51...53C,2000PhR...337..237C,2002ApJ...580..606H,2002MNRAS.331...23E,2009A&A...505..625G,2019MNRAS.484.1456E}) investigated the dependence of chaoticity in gravitational $N-$body systems as function of the number of particle $N$ in terms of the largest Lyapunov exponent (i.e. the rate of exponential divergence of initially nearby trajectories, \citealt{lichtenberg2013regular}). In particular, Kandrup and collaborators (\citealt{1996kandrup,2001PhRvE..64e6209K,Sideris_2002,kandrupvass,Kandrup_2003,kandrupvasssideris,Sideris_2004,2004CeMDA..88....1K}) studied how the collective properties of ensembles of trajectories relate to their average maximal Lyapunov exponents when propagated in both static non-integrable or periodically pulsated galactic potentials, as well as frozen $N-$body models (i.e. systems composed by $N$ fixed particles sampling a given continuum density $\rho(r)$ and exerting a Newtonian potential, possibly softened). The main results of these studies (see e.g. \citealt{2005NYASA1045....3M} for an almost complete review) can be summarized as
\begin{enumerate}
    \item The time scale of the perturbation growth of individual stellar orbits decreases as $N^{-1}$, as opposed to the $N^{1/3}$ scaling proposed by \cite{GurzadianSavvidy}.
    \item For sufficiently large values of $N$, the r.m.s. of the spatial position of initially localized orbit ensembles evolves as
\begin{equation}
\langle r\rangle \propto C(N)\exp(t/t_{\rm c}),
\end{equation}
where the factor $C$ is a decreasing function of the number of degrees of freedom (i.e. number of particles).
\item The point above essentially implies that orbits in $N$ body systems approach their counterparts in smooth potentials as $N$ increases. This established it is not necessarily true that the maximal Lyapunov exponent of an orbit ensemble decreases with increasing $N$.
\end{enumerate}
More recently, the direct $N-$body integrations of stable and unstable self consistent models by \cite{Di_Cintio_2019_revisited,2020MNRAS.494.1027D} evidenced that the largest Lyapunov exponent of the $N-$body Hamiltonian indeed scales as $N^{-\alpha}$ with $1/3<\alpha<1/2$, compatible with the \cite{GurzadianSavvidy} prediction convoluted with a $N^{1/2}$ shot noise related to the sampling of particles form a given model density profile. Despite this, the largest Lyapunov exponents of individual orbits appear to have little to no dependence on $N$ (at least in the explored range $10^3\leq N\leq 1.5\times 10^5$), while at fixed $N$ a strong dependence on both the specific orbital energy and angular momentum is evident, in good agreement with what other authors conjectured using tracer particles in external potentials.\\
\indent In the context of VR it remains to be determined how important is the role of discreteness effects associated to local chaotic dynamics (often dubbed micro-chaos) with respect to the complex dynamics arising from the rapidly changing bulk potential (i.e. macro-chaos). In this preliminary work we explore this matter, building on an idea sketched in \cite{2005NYASA1045....3M}, namely propagating a large number of tracer particles in a time-dependent potential reflecting that of a numerical simulation of cold collapse, using the numerical code introduced in \cite{Pasquato_2020} (see also \citealt{Di_Cintio_2019}) that allows one to treat the discreteness effects by means of a friction and diffusion process. In practice, the dynamics of the tracer particles is formulated in terms of Langevin-like equations of the form
\begin{equation}\label{langeq}
\ddot{\mathbf{r}}=-\nabla\Phi(\mathbf{r},t)-\eta\mathbf{v}+\delta\mathbf{f}(\mathbf{r});\quad \mathbf{v}=\dot{\mathbf{r}},
\end{equation}
where $\Phi$ is the (possibly time dependent) potential of the model, $\eta$ is a Dynamical friction coefficient (\citealt{chandra}) and $\delta\mathbf{f}$ a fluctuating stochastic force (per unit mass), whose distribution and typical intensity depends on the specific model at hand. The dependence on the number of particles of the underlying $N-$body model is therefore parametrized (see below) inside $\eta$ and $\delta f$. By doing so, one can probe the behaviour of tracer orbits  in systems with arbitrary large values of $N$.\\
\indent The rest of the paper is structured as follows, in Sect. \ref{models1} we discuss in detail the models and their time-dependent properties and introduce the numerical scheme used to integrate Eq. (\ref{langeq}). In Sect. \ref{experiments} we present and discuss the different experiments of violent relaxing systems and explore the interplay between micro and macro chaos. Finally, Sect. \ref{remarks} summarizes and presents the astrophysical implications. 
\section{Methods and numerical experiments}\label{models1}
We study the dynamics dictated by Eq. (\ref{langeq}) in three types of time-dependent potentials where
\begin{enumerate}
\item the whole parent density profile contracts and expands self-similarly with the damped oscillations of its scale radius $r_c$.
\item the inner density slope below a control radius smoothly converges to its final radius  
\item both the scale radius and density slope are varied with the latter protocols.
\end{enumerate}
\subsection{Models}
\begin{figure}
	\centering
	\includegraphics[width=\columnwidth]{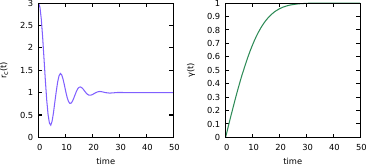}
	\caption{Evolution of the scale radius $r_c$ (left panel) and the logarithmic density slope (right panel)}
	\label{models}
\end{figure}

In all cases the density profile of the models is chosen from a the family of the so-called $\gamma-$models (\citealt{1993MNRAS.265..250D,1994AJ....107..634T})  
\begin{equation}\label{dehnen}
\rho(r)=\frac{3-\gamma}{4\pi}\frac{Mr_c}{r^\gamma(r+r_c)^{4-\gamma}},
\end{equation}
with total mass $M$, scale and logarithmic density slope $\gamma$. The gravitational associated potential reads
\begin{eqnarray}\label{phidehnen}
 \Phi(r)&=&-\frac{GM}{(2-\gamma)r_c}\left[1-\left(\frac{r}{r+r_c}\right)^{2-\gamma}\right]\quad {\rm for}\quad\gamma\neq 2;\nonumber \\
 \Phi(r)&=&\frac{GM}{r_c}\ln\frac{r}{r+r_c} \quad {\rm for}\quad\gamma=2.
 \end{eqnarray}
In the numerical experiments mimicking a self-similar dissipationless gravitational collapse, the density profile (and hence the corresponding potential) evolve self-similarly as the scale radius $r_c(t)$ undergoes damped oscillations with frequency $\omega$ around its (final) asymptotic value $r_{c,a}$ in the form given by
\begin{equation}
r_c(t)=2r_{c,a}\exp(-t^2/\tau_d^2)B_0(\omega t)+r_{c,a},
\end{equation}\label{damping}
where $\tau_d$ is the damping time and $B_0(x)$ is the Bessel function of the first kind of order 0. With such a choice, the time dependent scale radius $r_c(t)$ evolves with damped oscillations in a similar fashion as what happens during the virialization process to the so-called gravitational radius 
\begin{equation}
R_G(t)=-\frac{3}{5}\frac{GM^2}{U(t)},
\end{equation}
often used a a diagnostic quantity in $N-$body simulations of gravitational collapse (e.g. see \citealt{2012MNRAS.423.1610S} and references therein).\\
\indent When modeling a collapse affecting only the inner part of the system (say below $r_c$) we allow to smoothly vary the central logarithmic slope $\gamma$ from an initial value $\gamma_0$ to its asymptotic value $\gamma_a$ for fixed of damped $r_c$ as
\begin{equation}\label{collapse}
\gamma(t)=(\gamma_a-\gamma_0){\rm Erf}(t/\tau_d)+\gamma_0,
\end{equation}
where ${\rm Erf}(x)$ is the standard error function. In Fig. \ref{models} we show the time dependence of $r_c$ and $\gamma$ for the choice of simulation parameters discussed in the following, namely $\tau_d=14$, $\omega=0.85$, $\gamma_0=0$, $\gamma_a=1$ and $r_{c,a}=1$.\\
\indent Each tracer particle $m_t$ moving in $\Phi(r,t)$ is also subjected to friction and noise parametrized by the drag coefficient $\eta$ and the force fluctuation distribution $\mathcal{F}$. We implement the position and velocity dependent dynamical friction as
\begin{equation}\label{df}
\eta(r,v)=16\pi^2 G^2\rho(r)(m+m_t)\ln\Lambda\frac{\Psi(v)}{v^3}
\end{equation}
$\ln\Lambda$ is the Coulomb logarithm, $v=||\mathbf{v}||$, and
\begin{equation}\label{velfunct}
\Psi(v)=\int_0^{v}f(v^\prime)v^{\prime 2}{\rm d}v^\prime,
\end{equation}
is the fractional velocity volume function. In principle the (possibly time dependent) velocity distribution $f(v)$ is unknown, we adopt the usual local Maxwell-Boltzmann approximation, so that $\Psi$ then reads (see \citealt{2010AIPC.1242..117C}) as
\begin{equation}
\Psi(v)=(4\pi)^{-1}\left[{\rm Erf}\left(\frac{v}{\sqrt{2}\sigma}\right)-\sqrt{\frac{2}{\pi}}\frac{v}{\sigma}\exp\left(-\frac{v^2}{2\sigma^2}\right)\right],
\end{equation}
where $\sigma$ is the local velocity dispersion. For the case of the $\gamma-$models with density given in Equation (\ref{dehnen}) it reads
\begin{equation}
\sigma^2(r)=GMr^\gamma(r+r_c)^{4-\gamma}\int_r^\infty\frac{r^{\prime 1-2\gamma}{\rm d}r^\prime}{(r^\prime+r_c)^{7-2\gamma}}.
\end{equation}
The definition of Coulomb logarithm of the maximum to minimum impact parameter appearing in Eq. (\ref{df}) is somewhat arbitrary, throughout this work we adopt the time dependent analog of the widely used (see e.g. \citealt{binney}) 
\begin{equation}
\log\Lambda=\log\bigg[1+\bigg(\frac{r_cv^2}{Gm_t}\bigg)^2\bigg].
\end{equation}
It is known that the magnitude of the stochastic force (per unit mass) $\delta f$ in an infinite system of particles interacting via $1/r$ pair potential with homogeneous number density distribution with Poissonian fluctuations is described (see \citealt{chandra_vonnumann}) by the \cite{1919AnP...363..577H} distribution. The latter is known in integral form as
\begin{equation}\label{holtsmark}
\mathcal{F}(\delta f)=\frac{2}{\pi \delta f}\int_0^\infty\exp\left[-\alpha(s/\delta f)^{3/2}\right]s\sin(s){\rm d}s,
\end{equation}
where $s$ is a dimensionless auxiliary variable and the density dependent normalization factor $\alpha$ is
\begin{equation}
\alpha=\frac{4}{15}(2\pi G)^{3/2}m^{1/2}\rho.
\end{equation}
The models considered in the present work though, in principle, extended to infinity, have spherically symmetric density profiles with radially (and temporal) dependent density. We therefore accommodate the definition (\ref{holtsmark}) using the local values of $\rho$.
\subsection{Orbits integration}
Equation (\ref{langeq}) is integrated for each tracer particle using the so-called quasi-symplectic method, introduced in \cite{Mannella_2004}. For reasons of simplicity, but without loss of generality, we write it for the one dimensional case as
\begin{align}\label{mannella}
x^\prime&=x(t+\Delta t/2)=x(t)+\frac{\Delta t}{2}v(t)\nonumber\\
v(t+\Delta t)&=c_2\left[c_1v(t)+\Delta t \nabla\Phi(x^\prime)+d_1 \tilde{\delta f}(x^\prime) \right]\nonumber\\
x(t+\Delta t)&=x^\prime+\frac{\Delta t}{2}v(t+\Delta t).
\end{align}
In the expressions above $\Delta t$ is the fixed time-step (as a rule here we adopt $\Delta t= 10^{-3}t_{\rm dyn}$, with $t_{\rm dyn}\equiv\sqrt{r_c^3/GM}$ the crossing time of the asymptotic system, i.e. corresponding to the density profile at $t\rightarrow\infty$),  $\tilde{\delta f}$ is the normalized stochastic force (i.e., a dimensionless random variable sampled from the adimensionalized Eq. \ref{holtsmark}), and
\begin{equation}
c_1=1-\frac{\eta\Delta t}{2};\quad c_2=\frac{1}{1+\eta\Delta t/2};\quad d_1=\sqrt{2\zeta\eta\Delta t},
\end{equation}
where the effective energy $\zeta$, in the case of a delta correlated noise, is fixed by the standard deviation of $\mathcal{F}$ as
\begin{equation}\label{sd}
\langle \delta f(x,t) \delta f(x,t^\prime)\rangle=2\eta\zeta\delta(t-t^\prime).
\end{equation}
Since for the Holtsmark distribution (\ref{holtsmark}) the standard deviation, as well as all the other higher moments are singular, our numerical scheme uses instead $f_{\rm HM}/\sqrt{8\log{2}}$, where $f_{\rm HM}$ is the full width at half maximum of the truncated distribution. We note that, for vanishing $\eta$ and $\zeta$, Eqs. (\ref{mannella}) yield back the standard second order and symplectic Leapfrog method.
\begin{figure}
	\centering
	\includegraphics[width=\columnwidth]{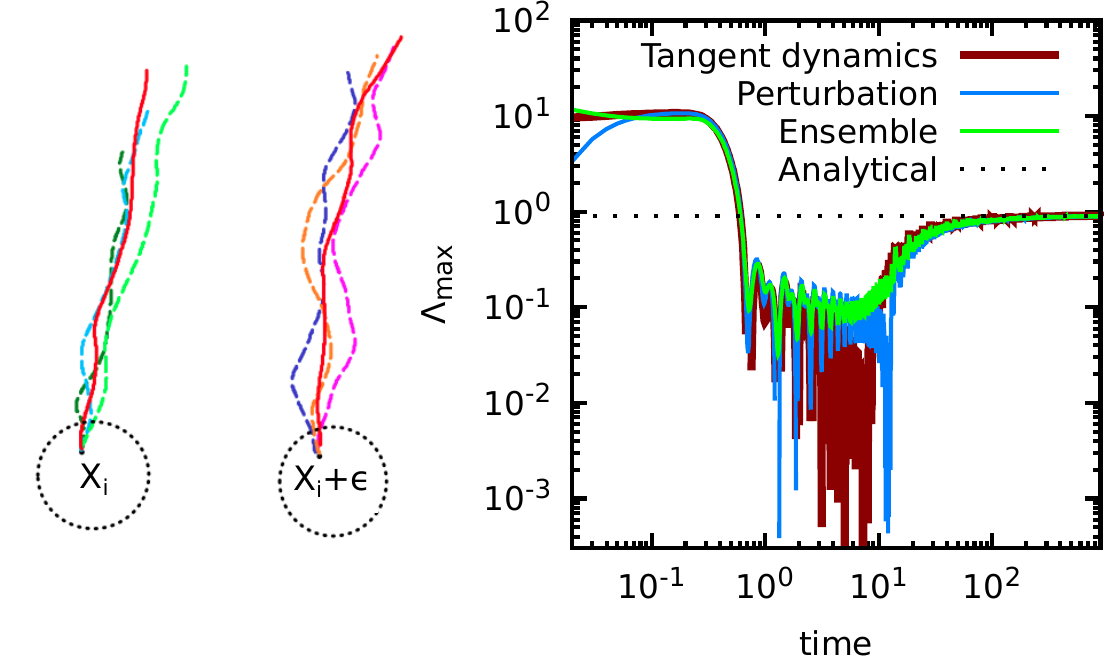}
	\caption{Sketch of the trajectories ensembles for the computation of the mean maximal Lyapunov exponent (left). Convergence of the maximal Lyapunov exponent $\Lambda_{\rm max}$ for the Lorenz system (right)}
	\label{fig:dotslyap}
\end{figure}
\subsection{Lyapunov exponents}
The largest Lyapunov exponent $\Lambda_{\rm max}$ is estimated in numerical simulations (see e.g. \cite{1976PhRvA..14.2338B,kantz2004nonlinear,Skokos_2009}) propagating up to a large time $t=L\Delta t$ as 
 \begin{equation}\label{lmax}
\Lambda_{\rm max}(t) =\frac{1}{L\Delta t}\sum_{k=1}^L\ln\frac{||\mathbf{W}(k\Delta t)||}{||\mathbf{W}_0||}~,
\end{equation}
where $\mathbf{W}$ is the tangent-space vector, that for a 3D $N-$body system is the $6N$-dimensional vector
\begin{equation}\label{tangent}
\mathbf{W}=(\mathbf{w}_i,\dot{\mathbf{w}}_i,...,\mathbf{w}_N,\dot{\mathbf{w}}_N).
\end{equation}
In the definitions above $(\mathbf{w}_i,\dot{\mathbf{w}}_i)$, are the tangent sub-space coordinates of the $i-$th degree of freedom (particle), and $||...||$ is the standard Euclidean norm, here applied in $\mathbb{R}^{6N}$ to the normalized $\mathbf{w}_i$s and $\dot{\mathbf{w}}_i$s. The vector $\mathbf{W}$ in initialized to $\mathbf{W}_0$ at time $t=0$. Technically speaking, the Lyapunov exponents are rigorously defined only in the infinite-time limit. In practical numerical simulations, we evaluate their approximate discrete definition (Eq. \ref{lmax}) them through sufficiently long integrations until convergence is numerically established. In the numerical experiments discussed hereafter we fixed a tolerance of the 0.001\% on the fluctuations of the time series.\\
\indent In systems where the tangent dynamics is not known or computationally unpractical (see e.g. \citealt{2025A&A...693A..53D} and the discussion therein) one typically substitutes to $\mathbf{W}$ the distance between the reference orbit and that starting from a slightly perturbed initial condition, given by
\begin{equation}\label{pert}
\Tilde{\mathbf{W}}=(\mathbf{r}_i-\mathbf{r}_i^\prime,\dot{\mathbf{r}}_i-\dot{\mathbf{r}}_i^\prime,...,\mathbf{r}_N-\mathbf{r}_N^\prime,\dot{\mathbf{r}}_N-\dot{\mathbf{r}}_N^\prime),
\end{equation}
where the primed quantities refer to the perturbed phase-space trajectory. Defining a numerical Lyapunov exponent for a $N=1$ dynamics governed by Equation (\ref{langeq}) is problematic for a twofold reason (see \citealt{10.1007/BFb0076835,Laffargue_2016,2024arXiv241107064B}), on one side due to the absence of a well defined tangent dynamics for the stochastic term, on the other due to the dependence of a given trajectory on the specific choice of random variables in implementing the noise protocol.\\
\indent Here we propose a simple empirical scheme\footnote{We recall that \citet{Kandrup_2003} also presented a scheme to evaluate $\lambda_{M}$ for models with a stochastic component, based on averages over orbit segments.} to estimate a proxy of the Lyapunov exponent of a particle orbit in a potential subjected to noise and friction, using multiple realization of the perturbed trajectories appearing in the definition ($\ref{pert}$). For each tracer particle's initial condition $\mathbf{X}_i$ we propagate $N_{\rm e}$ trajectories using different chains of pseudo-random numbers to sample the stochastic term, and another $N_{\rm e}$ trajectories starting from slightly perturbed initial condition $\mathbf{X}_i+\mathbf{\epsilon}$ with fixed $||\mathbf{\epsilon}||$ and independent chains of pseudo-random numbers, as sketched in the left panel of Fig. \ref{fig:dotslyap}. By doing so, we can therefore evaluate a set of $N_{\rm e}$ putative values of the largest Lyapunov exponent $\lambda_{*}$ whose root means square defines the value of the largest Lyapunov exponent of the wanted orbit defined by Eq. (\ref{langeq}) as
\begin{equation}
    \lambda^2_{M}  = \frac{1}{N_{e}}\sum_{i=1}^{N_{e}}(\lambda_{*,i}-\bar{\lambda}_{*})^2
\end{equation}
where $\bar{\lambda}_{*}$ is the mean value of ${\lambda}_{*,i}$. In this work, as a rule, we used $N_{\rm e}=20$ and $\epsilon=10^{-6}$.\\
\indent Our model considers effectively a dissipationless scenario, with semi-analytical noise-friction processes introduced to mimic discreteness effects inherent in $N$-body systems. Hence, our total sum of Lyapunov exponents indeed remains close to zero by construction. A slight physical or numerical dissipation, would theoretically make this sum negative and thus influence longer-run dynamics. This is only evident in prohibitively long numerical integrations, often exceeding the time scale of the physical process of interest.\\
\indent In order to verify that in absence of noise the procedure introduced above gives an acceptable estimate of the largest Lyapunov exponent, we have applied it to a simple chaotic model for which the variational equations necessary to integrate the tangent dynamics are simple and the largest Lyapunov exponent $\Lambda_{\rm max}$ was accurately estimated  (\citealt{1985PhyD...16..285W}). We considered the paradigmatic \cite{DeterministicNonperiodicFlow} system, defined by the set of three first order ODEs
\begin{align}
\dot{x}&=\varsigma(y-x)\\\nonumber
\dot{y}&=x(\varrho-z)-y\\\nonumber
\dot{z}&=xy-\beta z.
\end{align}
For the specific choice of model parameters $\varsigma=10$, $\varrho=28$ and $\beta=8/3$, it is known that $\Lambda_{\rm max}\approx 0.906$. In the right panel of Fig. \ref{fig:dotslyap}, as an example, we show the convergence of the numerical estimation of the largest Lyapunov using an integration of the variational (tangent) dynamics, a single perturbed trajectory, and an ensemble average over 10 nearby initial conditions. remarkably, the value of $\Lambda_{\rm max}$, marked in figure by the horizontal dashed line is rather well approached in all cases.  
\begin{figure}
	\centering
    \includegraphics[width=0.9\columnwidth]{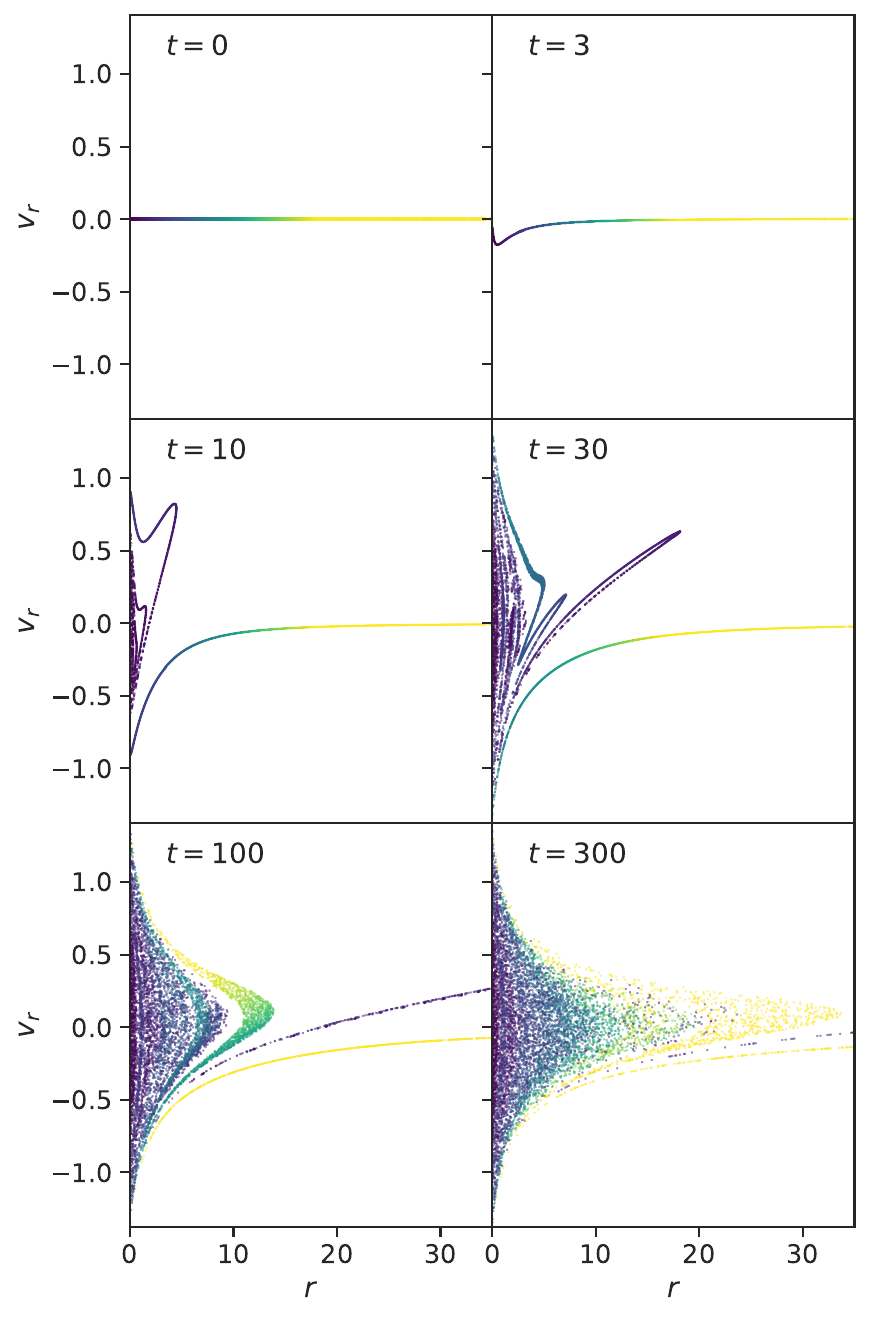}
	\caption{Radial phase-space section for an ensemble of tracers in a time dependent potential simulation the self-similar collapse of a $\gamma=1$ model in the continuum limit. The color coding indicates the initial radial position of each particle.}
	\label{phase}
\end{figure}
\begin{figure}
	\centering
    \includegraphics[width=0.9\columnwidth]{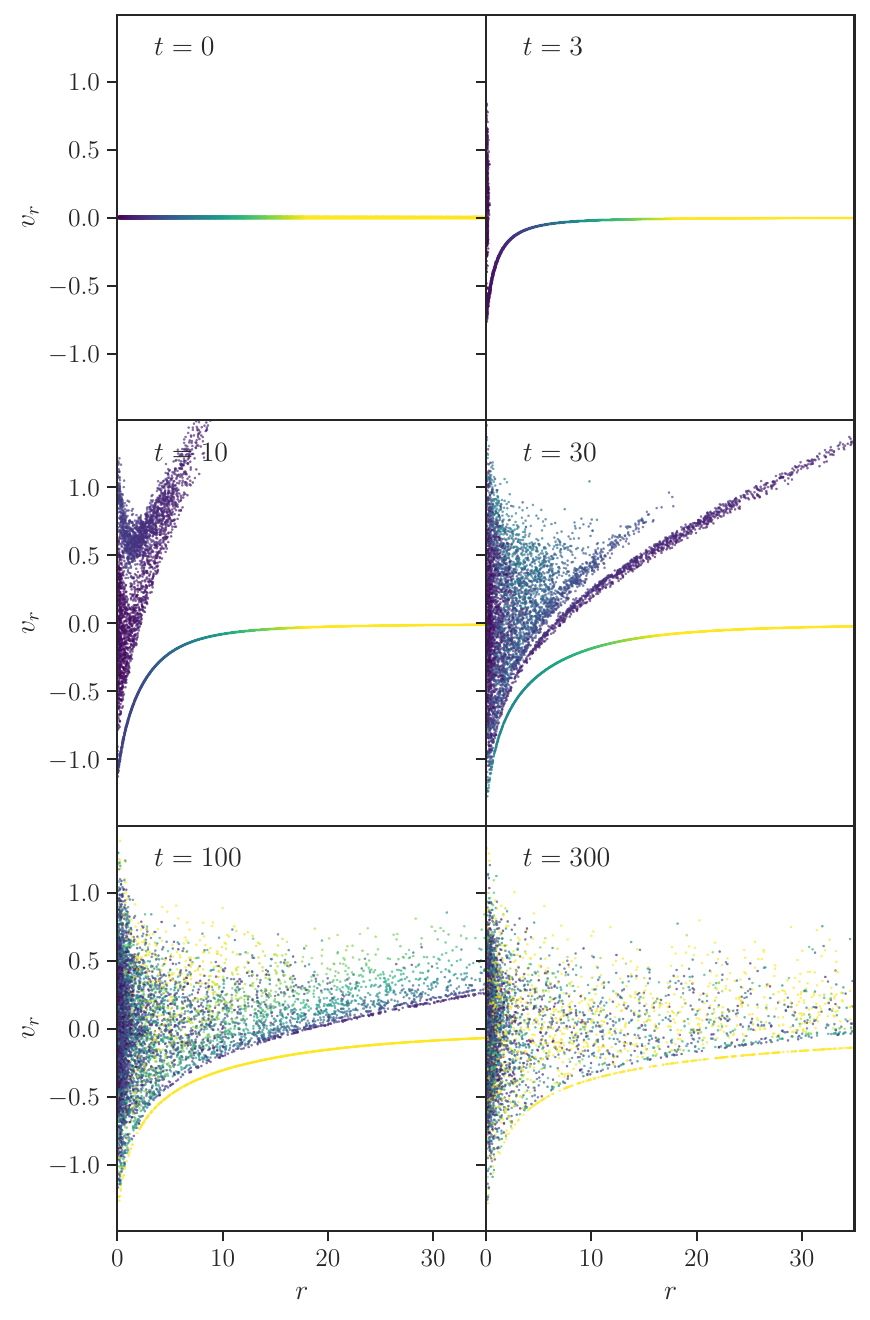}
	\caption{Same as in Fig. \ref{phase} but for a run including discreteness effects corresponding to a $N=10^4$ system. Again, The color coding indicates the initial radial position of each particle.}
	\label{phase1e4}
\end{figure}
\begin{figure}
	\centering
    \includegraphics[width=0.9\columnwidth]{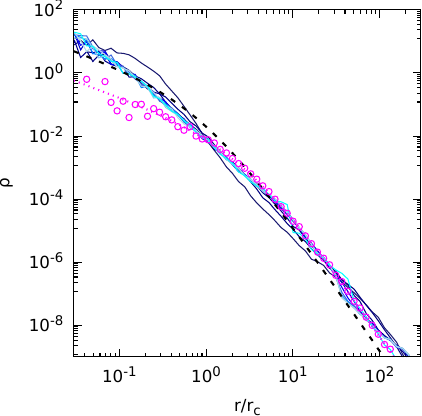}
	\caption{Density profile at $t=300$ for the $N_{\rm t}=10^4$ tracers in models with effective $N$ increasing from $10^4$ to $10^{13}$, corresponding to increasingly lighter tones of blue (solid lines). Initial and final density profiles of the background model (purple and black dashed lines, respectively) and initial density profile of the tracers (points).}
	\label{rho}
\end{figure}
\subsection{Emittance and dynamical entropy}
Defining a dynamical entropy for a system of independent particles with dynamics governed by Eq. (\ref{langeq}) is rather subtle, due to the fact that for the particular case study of the present work the disorder and relaxation are induced by and external potential and stochastic process mimicking a real collective $N-$dynamics. Here, following \cite{Di_Cintio_casetti} (see also, \citealt{2005NYASA1045...12K} and references therein), we have evaluated the temporal evolution of the so called emittance, a quantity introduced in the context of charged
particles beams (see e.g. \citealt{1991JAP....70.1919R}) and defined by
\begin{equation}\label{emit}
    \varepsilon = \left ( \varepsilon_x \varepsilon_y \varepsilon_z  \right )^{1/3} \text{,}  \quad \varepsilon_i=\sqrt{\left\langle r_i^2\right\rangle\left\langle v_i^2\right\rangle-\left\langle r_i v_i\right\rangle^2}
\end{equation}
where $\left \langle \cdot  \right \rangle$ denotes the ensemble averages.
In the dynamics of a charged particle beam in an accelerator, the emittance quantifies the degree to which the individual particles depart from a idealized beam without space-charge effect. That is, a small emittance implies that particles are tightly ''packed" in phase-space. Conversely, increasingly larger values of $\varepsilon$ indicate that particles are more spread out over the accessible phase-space.\\
\indent We recall that the definition (\ref{emit}) may be also thought as the determinant of the variance-covariance matrix of the beam's phase-space coordinates as
\begin{equation}
\varepsilon = \sqrt{\left|\begin{array}{cc}
\langle x \cdot x\rangle & \left\langle x \cdot x^{\prime}\right\rangle \\
\left\langle x \cdot x^{\prime}\right\rangle & \left\langle x^{\prime} \cdot x^{\prime}\right\rangle
\end{array}\right|} \quad,
\end{equation}
from which it becomes evident that $\varepsilon$ quantifies an effective (hyper-)surface enclosing the effective phase-space (hyper-)volume occupied by the system, in terms of its second order statistics.\\
\indent We note that \cite{2019ApJ...872...20B,2019ApJ...870..128B} investigated the role of discreteness effects in effective $N-$body models using the discrete formulation of the Henon, Lynden-Bell \& Tremaine entropy (hereafter HLBT, see \citealt{1986MNRAS.219..285T})  defined for a particle-based model as 
\begin{equation}\label{shat}
\hat{S}=\frac{1}{N}\sum_{i=1}^N\ln D_i^6+\ln\left[\frac{\pi^3}{6}(N-1)\right]+\gamma_{\rm EM},
\end{equation}
where
\begin{equation}\label{dist}
D_i=\sqrt{(\mathbf{r}_i-\mathbf{r}_{n,i})^2+(\mathbf{v}_i-\mathbf{v}_{n,i})^2}
\end{equation}
and $\gamma_{EM}\approx 0.57722$ is the Euler-Mascheroni constant. In the definition above $\mathbf{r}_{n,i}$ and $\mathbf{v}_{n,i}$ are the instantaneous position and velocity of the nearest phase-space neighbour of particle $i$ \citep{2008arXiv0810.5302L}.  In a self consistent $N-$body simulation of a VR process $\hat{S}$ is expected to evolve towards a nearly invariant value. However, the same is also true for $N$ independent particles starting from particular initial conditions and evolved in either regular or chaotic, static potentials, as shown in \cite{2019ApJ...870..128B}. The same is also true for the claimed $N^{1/6}$ scaling of the relaxation time scale associated with the growth of $\hat{S}$ for increasingly larger ensembles of tracer trajectories integrated in the same fixed external potential. This stems from the definition of $D_i$ in Eq. \ref{dist} entering Eq. (\ref{shat}), that manifestly yields an average interparticle distance in $\mathbb{R}^6$. The expression for the average interparticle distance for an ensemble of $N$ particles in a volume $V \in \mathbb{R}^6$ reads as:
\begin{equation}
    \langle d \rangle_{\mathbb{R}^6} = \frac{\Gamma \left(\frac{1}{6}\right)
   }{6^{5/6} \sqrt{\pi }} \sqrt[6]{\frac{V}{N-1}},
\end{equation}
from which the timescale dependence ${\propto} N^{-1/6}$ found in \citet{2019ApJ...870..128B}  naturally arises. We stress the fact that, our method potentially bears some limitations, as it is based on the propagation of tracer particles. However, in our case, the dynamics of said tracers become implicitly correlated by the local discreteness effects parametrized by $\eta$ and $\delta\mathbf{f}$. \\
\indent It is well known that the emittance as defined in Eq. (\ref{emit}) can be also thought as an entropy indicator (see e.g. \citealt{505881,1996PhRvE..54..830S}). In particular \cite{Di_Cintio_casetti} showed that $\varepsilon$ is constant for self-consistent equilibrium $N-$body models, provided that $N$ is sufficiently large, while it increases exponentially for initially localized ensembles of particles, in both frozen and active $N-$body gravitational models. A similar exponential emittance growth has been also reported by \cite{2005NYASA1045...12K} for the simulations of relaxing charged particle beams where the time evolution of $\varepsilon$ is rather well fitted by
\begin{equation}\label{emitfit}
\varepsilon(t)=C\sqrt{\frac{t}{N}}\exp(t\langle\lambda\rangle),
\end{equation}
where $C$ is a dimensional constant and $\langle\lambda\rangle$ is the ensemble averaged largest Lyapunov exponent. We note that another important definition of dynamical entropy associated to the (spectrum of) trajectories' Lyapunov exponents is the Kolmogorov-Sinai (KS, \citealt{kolmogorov0,kolmogorov,sinai}) entropy $\mathcal{S}_{\rm KS}$ (see also \citealt{lichtenberg2013regular}). Given a discretization of time and phase-space, $S_{\rm KS}$ quantifies the probability that a trajectory occupies a cell $j$ at a time $t_{n+1}$, conditioned on its history up to $t_n$. \cite{1977RuMaS..32...55P} proved that the KS entropy of an orbit is always upper bounded by the sum of its positive Lyapunov exponents
\begin{equation}\label{kse}
\mathcal{S}^+_{\rm KS}=\sum_{\lambda_i>0}\lambda_i.
\end{equation}
\cite{2025A&A...693A..53D} used $\mathcal{S}^+_{\rm KS}$ in addition to the largest Lyapunov exponent as an indicator of chaos to investigate the effect of increasing relativistic perturbations in some configurations of the gravitational three body problem, finding that $\mathcal{S}^+_{\rm KS}$ bears a different trend with the strength of such perturbations than that of $\Lambda_{\rm max}$. In this work, however, a clear definition of Lyapunov spectrum is not available because we are dealing with particle dynamics with stochastic terms, so we therefore resort to the numerical evaluation of dynamical entropies not involving the other $\lambda$s. 
\begin{figure}
	\centering
	\includegraphics[width=0.8\columnwidth]{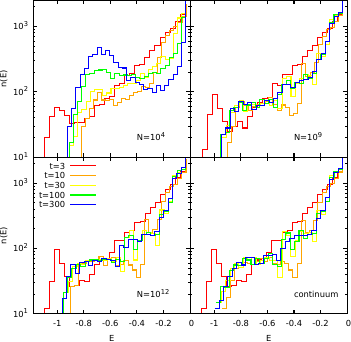}
	\caption{Evolution of the number energy distribution $n(E)$ for $N=10^4$, $10^9$, $10^{12}$ and the continuum limit of a self-similar collapse.}
	\label{figne}
\end{figure}
\begin{figure*}
	\centering
	\includegraphics[width=\textwidth]{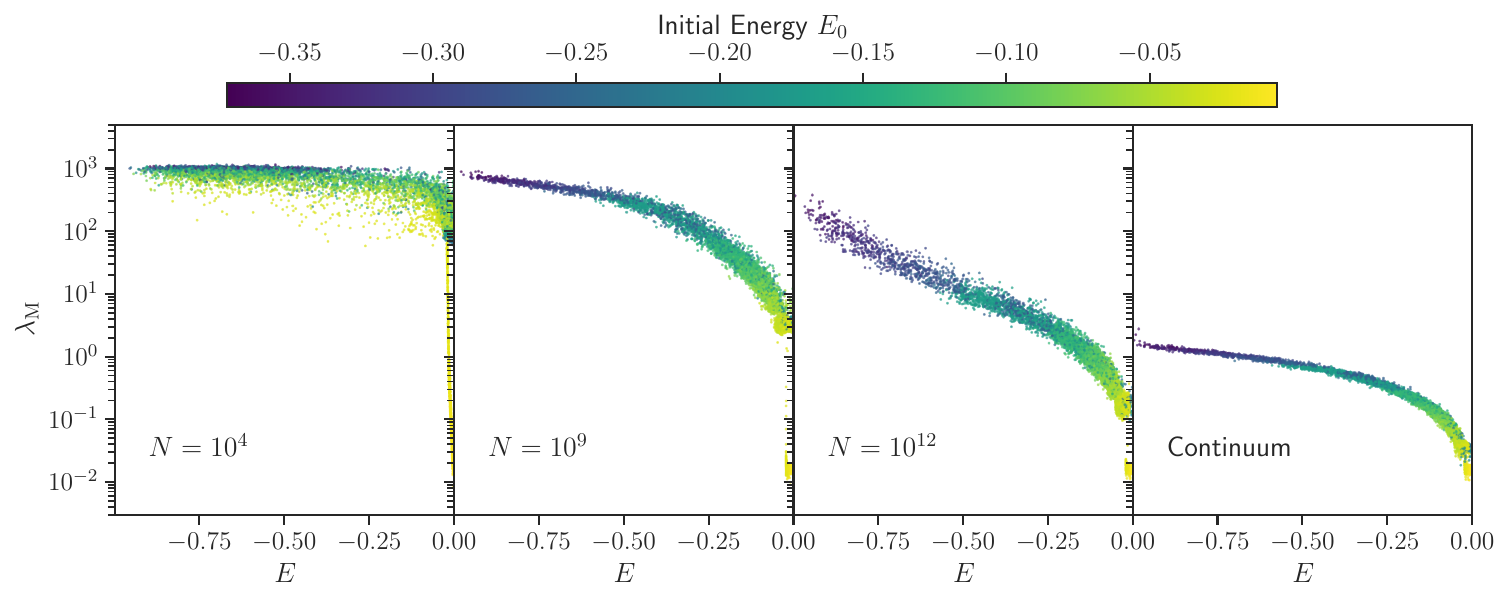}
	\caption{Largest Lyapunov exponent $\lambda_{\rm M}$ as function of the final orbital energy for $10^4$ particles propagating in effective self-similar collapsing $\gamma=1$ system with $N=10^4$, $10^7$, $10^{12}$ and infinite. The color coding marks the initial value of particles specific energies $E_0$.}
	\label{lyap_e0}
\end{figure*}
\begin{figure*}
	\centering
	\includegraphics[width=\textwidth]{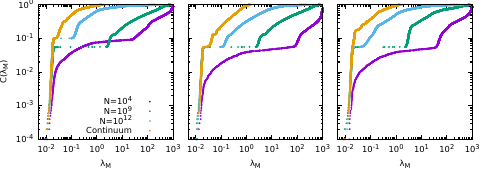}
	\caption{Cumulative distribution of the tracer orbits larger Lyapunov exponents for, from left to right, a self similar collapse, a steepening of the inner density profile and a collapse with strengthening of the central density cusp. The different sets of points mark the $N=10^4$ (purple), $N=10^9$ (green), $N=10^{12}$ (teal) and the continuum limit (orange)}	\label{lyap_dist}
\end{figure*}
\begin{figure}
	\centering
	\includegraphics[width=\columnwidth]{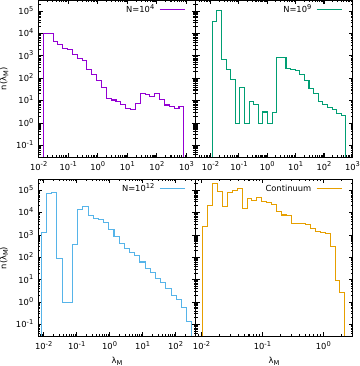}
	\caption{Distribution of maximal Lyapunov exponents for the tracer orbits in a self similar collapse model, corresponding to the cumulative distribution of the left panel in Fig. \ref{lyap_dist}}
	\label{lyap_nl}
\end{figure}
\begin{figure*}
	\centering
	\includegraphics[width=\textwidth]{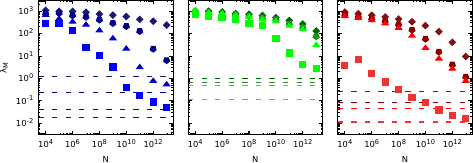}
	\caption{$N$-scaling of the mean maximal Lyapunov exponent for ensembles of tracers starting with different initial velocities, for the same experiments as in Fig. \ref{lyap_N}. The horizontal dashed lines mark the values attained for the continuum model without force fluctuations and friction.}
	\label{lyap_e}
\end{figure*}
\begin{figure*}
	\centering
	\includegraphics[width=\textwidth]{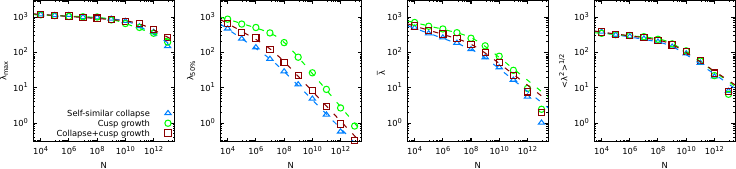}
	\caption{From left to right, for an ensemble of tracers covering the whole accessible phase-space, $N$-scaling of the largest Lyapunov exponent  $\lambda_{\rm max}$, median Lyapunov exponent $\lambda_{50\%}$, and mean value $\Bar{\lambda}$ and root mean square of the distribution $\langle\lambda^2\rangle^{1/2}$. The blue triangles, green circles and red square correspond to numerical experiments simulating a self-similar collapse, a steepening of the central density cusp ($\gamma$ from 0 to 1) and a collapse with cusp steepening, respectively.}
	\label{lyap_N}
\end{figure*}
\section{Numerical experiments and results}\label{experiments}
\subsection{Structural properties of the cold collapse}
We integrated a wide set of orbits in time dependent potentials with $N-$dependent noise and friction coefficients, simulating spherical dissipationless collapses.\\ 
\indent In the simulations discussed here, we restrict ourselves to the cases where the initial conditions of $N_{\rm t}=10^4$ tracer particles are characterized by zero velocity and sample uniformly the density profile\footnote{The density of tracer particles divided by the effective $\rho$ of the background system is constant over all radii.} corresponding to the embedding potential at $t=0$. By doing so, we are considering a purely cold gravitational collapse (i.e. total initial kinetic energy $K_0=0$), that is the typical set up for $N-$body simulations to undergo a VR process (\citealt{1982MNRAS.201..939V,1991MNRAS.250...54L,2005A&A...433...57T}). As a rule, all simulations are extended up to $300t_{\rm dyn}$, so that the fluctuations of the numerically evaluated Lyapunov indicators fall within the requested tolerance window.  In Fig. \ref{phase} we show the radial phase-space sections of $N_{\rm t}=10^4$ tracer particles (i.e. radial coordinate $r$ against radial velocity $V_r$) at times 0, 3, 10, 30, 100 and 300 in units of $t_{\rm dyn}$ in a self-similar collapse of a $\gamma=1$ model without noise and friction (i.e. continuum limit). In this specific case the virial oscillation frequency $\omega=0.85$, the damping time $\tau_d=13t_{\rm dyn}$, and initial and final scale radii $r_c=3$ and 1, respectively. Remarkably, the distribution of points appears to be strikingly similar to that observed in self-consistent collapse simulations (see analogous plots in \citealt{1991MNRAS.250...54L,2007ApJ...660..256N,2013MNRAS.431.3177D}). The points colour coding marks the radial\footnote{Since in this particular numerical experiment the initial kinetic energies were set to 0, the total particle initial energy (i.e. potential only) is directly proportional to the initial radial position.} positions at $t=0$ revealing that, even in absence of discreteness effects, a rapidly varying potential is enough to induce mixing and hence energy relaxation. In Fig. \ref{phase1e4} we show the same distribution of phase-space coordinates for the case where the parent potential is perturbed by discreteness effects via friction and noise, corresponding to a $N=10^4$ $N-$body system. In this case, already at $t=10$, we notice a more efficient mixing, represented by a stronger dispersion of points of the same colour, due to the action of the effective granularity. This is particularly evident in the final state at $t=300$, that appears structurally similar to its parent continuum limit shown in Fig. \ref{phase}, but with individual tracer particle phase-space positions almost entirely decorrelated from their state at $t=0$.\\
\indent In Fig. \ref{rho} We show as function of $N$ (represented by the increasingly lighter shades of blue) the final radial density distribution at $t=300$ of the $N_{\rm t}$ tracers for the case of a $\gamma=1$ self-similar collapse with $r_c$ starting from 3 (magenta dashed line) and relaxing to 1 (black dashed line) with damped oscillations. With the notable exception of the discreteness effects-dominated $N=10^4$ case (corresponding to the dark blue line bearing a evident core-halo structure), all density profiles approach that of the asymptotic relaxed system, with some discrepancies related to the fixed number of tracer particles that, as a rule, we fixed to $10^4$. When the evolution of the external potential is characterized by values of $\omega^{-1}$ and $\tau_c$ much larger than a typical crossing time (i.e. the evolution is ''adiabatic"), the final density profile of the tracer distribution is expected to match to a higher degree that of the supporting external potential, as it happens in adiabatic squeezing techniques used to generate approximate equilibrium triaxial models starting from spherical $N-$body realizations with ergodic distribution functions (see \citealt{2001ApJ...549..862H}).\\
\indent Another quantity used to probe the mixing properties of phase-space is the differential energy distribution $n(E)$ defined from the phase-space distribution function $f$ (see e.g. \citealt{binney,2021A&A...653A.140B}) as
\begin{equation}
n(E)=\iint_\Pi f(\mathbf{r},\mathbf{v})\delta\big[E+{v^2}/{2}-\Phi(\mathbf{r})\big]{\rm d}\mathbf{r}{\rm d}\mathbf{v},
\end{equation}
where $\delta$ is the usual Dirac delta and $\Pi$ is the phase-space volume occupied by the system. Figure \ref{figne} shows the evolution of $n(E)$ in the self-similar collapse for the $N=10^4$, $10^9$, $10^{12}$ and the continuum limit. Consistently with the density profile $\rho$, tracer particles in models with larger values of $N$ do evolve towards nearly indistinguishable energy distributions, dictated by the VR. The lowest $N=10^4$ case, where the discreteness effects parametrized by $\eta$ and $\mathcal{F}$ are the largest, stands out for showing a markedly bimodal $n(E)$. Interestingly, a similar form of the differential energy distribution has been also reported for $N-$body experiments of dissipationless collapses (see e.g. \citealt{2013MNRAS.431.3177D}) for the cases with larger initial virial ratios (typically around $0.4-0.5$), generally associated with a core-halo relaxed density distribution $\rho$. Remarkably, despite the fact that the simulations performed in this work were initialized with zero initial ''temperature", the larger the discreteness effects the closer is the dynamics of tracer particles to those of system collapsing with a higher initial virial ratio. Moreover, this is also coherent with what found for $N-$body simulations as function of their particle resolution at fixed type of initial condition (e.g. see \citealt{2024A&A...683A.254D}).  
\subsection{Lyapunov exponents}
It is well known that even in low dimensional models and for simple non-integrable potentials, the estimated values of the Lyapunov exponents might depend strongly on the orbital energy and even be different for initial conditions starting on the same energy surface but different phase-space coordinates. This is evident in Fig. \ref{lyap_e0} where we show for the same models of Fig. \ref{figne} the tracer particle largest Lyapunov exponent $\lambda_{\rm M}$ as function of its orbital asymptotic energy $E$ and (cfr. the colour coding) its initial energy $E_0$. In general and consistently with the direct $N-$body calculations of \cite{Di_Cintio_2019_revisited} (see Fig. 10 therein), orbits that are localized in central regions and are therefore associated with lower energies show larger values of $\lambda_{\rm M}$ (i.e. are more chaotic) than those of weakly bound particles (i.e. $E\approx 0^-$). larger discreteness effects (i.e. lower $N$) induce a broader spread of $\lambda_{\rm M}$ at fixed final $E$ while also, as expected, triggering a more efficient energy mixing, possibly even after the VR phase than models governed by collective (external) potentials.\\
\indent Individual values of the Lyapunov exponents are typically lower at given $E$ for increasing values of $N$, however, moving in the direction of the continuum limit, those associated to weakly bound orbits have a stronger dependence with $N$, while between $N=10^4$ and $10^9$, where discreteness effects are still present but not dominant, the larger values of $\lambda_{\rm M}$ associated to orbits with $E\approx -1$ are substantially unchanged. This is also evident from the cumulative distribution  $C(\lambda_{\rm M})$ (i.e. ordered plot) over the whole ensemble of test particle orbits, shown in Fig. \ref{lyap_dist} for system sizes $N=10^4$ (purple), $10^9$ (green), $10^{12}$ (teal) and the continuum limit (orange) and, from left to right, a self-similar collapse of a $\gamma=1$ model, a central cusp growth, and a collapse with steepening of the central cusp from $\gamma_0=0$ to $\gamma_a=1$. In all cases, we find that for $N\gtrsim 10^8$ the normalized cumulative distribution of Lyapunov exponents converges to that of the continuum limit model below a critical value $\lambda_{\rm M}$ above which the distributions differ significantly having different slopes. Curiously, for intermediate and large $N$, $C(\lambda_{\rm M})$ has several gaps that narrow in the continuum limit. We verified that this peculiar feature is not an artifact of the specific choice of initial conditions. The presence of said gaps, implies that the associated differential distribution of $n(\lambda_{\rm M})$ has a gapped and multi-modal structure as evidenced in Fig. \ref{lyap_nl} for the case of a self-similar collapse.\\
\indent Following \cite{Di_Cintio_2019_revisited} in their analysis of equilibrium models, we computed $\lambda_{\rm M}$ for reference orbits starting from the same initial conditions at fixed initial energy $E_0$, evolving in models with different values of $N$. Figure \ref{lyap_e} shows $\lambda_{\rm M}$ as a function of $N$ for $E_0=-0.35$ (diamonds), $-0.1$ (pentagons), $-0.05$ (triangles) and $-0.01$ (squares) in self-similar collapses with $\gamma=1$ (left), cusp steepening (middle) and collapse with cusp steepening (right). Consistently with what observed in Fig. \ref{lyap_e0}, at fixed $N$ larger (negative) initial energies reflect in more chaotic orbits, while particles initially less bound are typically associated with smaller values of their largest Lyapunov exponent. The latter scales with $N$ for the considered values of $E_0$ with markedly different trends, though all monotonically decreasing. Orbits starting with energies $E_0\rightarrow 0^-$ yield for large $N$ values of finite time Lyapunov exponents that are closer to their counterparts for the continuum model, where only the time-dependent collective potential mixes particle energies. The values of $\lambda_{\rm M}$ for this limit are indicated in figure by the horizontal dashed lines. We note that the Lyapunov indicators associated to particle energies of strongly bound orbits are far from their converged values in the continuum limit even for large and astrophysically relevant system sizes, which implies that orbits in the central regions of giant galaxies with $N\approx 10^{12}\div 10^{13}$ stars might as well be wildly chaotic, even in absence of significant deviations from the spherical symmetry.\\
\indent In order to quantify the dependence of the chaos indicator on the strength of the discreteness effects we have evaluated as function of $N$ the relevant moments of the distribution of largest Lyapunov exponents of the tracer orbits, namely the maximum $\lambda_{\rm max}$, the median $\lambda_{50\%}$, the mean $\bar{\lambda}$ and the root mean square $\langle\lambda^2\rangle^{1/2}$. In Fig. \ref{lyap_N} said quantities are reported for the self similar collapse (triangles), cusp steepening (circles) and collapse with cusp steepening (squares). In all cases the moments of $n(\lambda_{\rm M})$ have a broken power-law trends with $N$, nicely fitted (cfr. the dashed lines in figure) by 
\begin{equation}\label{eqfit1}
\lambda(N)=c\frac{\sqrt{N_*}}{N^\alpha(N+N_*)^\beta},
\end{equation}
where $N_*$ is a scale system size and $c$ a dimensional quantity ranging from 0.5 to $\sim190$. We observe that the trends of maximum $\lambda_{\rm max}$ and the root mean square $\langle\lambda^2\rangle^{1/2}$ bear no significant  difference along the three classes of numerical experiments, and the distribution of values at fixed $N$ has little to no scattering. In the other two cases, the median and the mean, the trend for the collapses with cusp steepening is systematically intermediate between that for the self similar collapse and the central cusp steepening. The values of $\alpha$, $\beta$ and $N_*$ are summarized in Tab.~(\ref{tab:2x2}) below.\\
\indent The scale size lies in the range $1.2\times 10^7 \lesssim N_*\lesssim 8\times 10^{10}$, 
\begin{table}[h]
 \caption{Fit parameters for Eq. (\ref{eqfit1}).}
    \centering
    \begin{tabular}{lccc}
       Indicator & $\alpha$ & $\beta$ & $N_*$ \vspace{1pt}\\\hline\hline
       \multicolumn{4}{c}{Self-similar collapse} \\ \hline
       $\lambda_{\rm max}$ & 0.023 & 0.213  & $1.86\times 10^{10}$\\
       $\lambda_{50\%}$ & 0.274 & 0.131 & $1.75\times 10^7$\\ 
       $\bar{\lambda}$ & 0.142 & 0.178 & $2.36\times 10^8$\\ 
       $\langle\lambda^2\rangle^{1/2}$ & 0.065 & 0.213 & $8.29\times 10^8$\\\hline
       \multicolumn{4}{c}{Cusp growth} \\ \hline
       $\lambda_{\rm max}$ & 0.016 & 0.133  & $9.12\times 10^{8}$\\
       $\lambda_{50\%}$ & 0.121 & 0.350 & $4.77\times 10^7$\\ 
       $\bar{\lambda}$ & 0.094 & 0.229 & $1.49\times 10^8$\\ 
       $\langle\lambda^2\rangle^{1/2}$ & 0.035 & 0.247 & $5.04\times 10^8$\\  \hline    
    \multicolumn{4}{c}{Collapse with cusp growth} \\ \hline
       $\lambda_{\rm max}$ & 0.029 & 0.182  & $7.96\times 10^{10}$\\
       $\lambda_{50\%}$ & 0.222 & 0.202 & $1.23\times 10^7$\\ 
       $\bar{\lambda}$ & 0.118 & 0.129 & $1.36\times 10^8$\\ 
       $\langle\lambda^2\rangle^{1/2}$ & 0.053 & 0.212 & $6.12\times 10^8$\\ \hline\hline
    \end{tabular}
    \label{tab:2x2}
\end{table}
that implies for all the indicators the transition between a rather flat slope for low $N$ and a $N^{1/6\div 1/3}$ trend takes place  at around $N\approx 10^9$. This could be interpreted as predominance of micro-chaos below $N_*$, while at $N>N_*$, the chaoticity of particle orbits is essentially due to the collective potential oscillations (coherent or incoherent) or possibly to the non-integrability of such potential itself for the case of static models (e.g. triaxial system, not explored here).
\begin{figure*}
	\centering
	\includegraphics[width=0.95\textwidth]{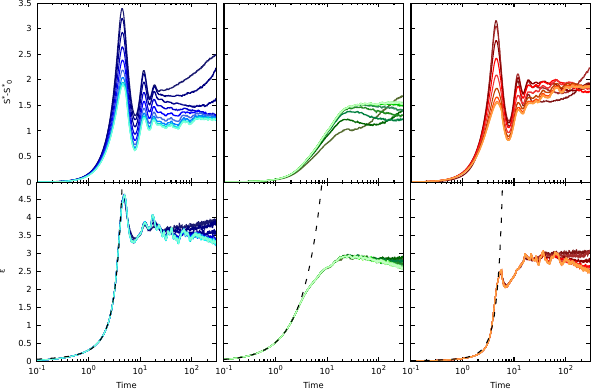}
	\caption{Modified HLBT entropy $S^*$ (top) and emittance $\varepsilon$ growth (bottom) for, from left to right, self-similar collapse, cusp growth and collapse with cusp growth. Increasingly lighter shades of blue/green/red mark increasingly larger values of $N$. The dashed lines mark the best fit growth trend of the exponential phase.}
	\label{emit_N}
\end{figure*}
\subsection{Dynamical entropy}
\begin{figure}
	\centering
	\includegraphics[width=0.9\columnwidth]{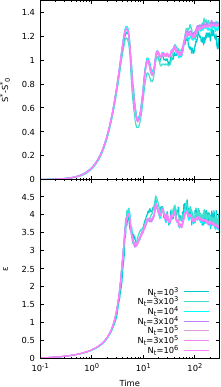}
	\caption{Modified HLBT entropy $S^*$ (top) and emittance $\varepsilon$ growth (bottom) for a collapse with cusp growth with $N=10^{11}$ and different numbers of tracer particles in the range $10^3\leq N_{\rm t}\leq 10^6$.}
	\label{emit_11}
\end{figure}

As we are dealing with a fixed number of tracer particles $N_{\rm t}<N$, we modified the definition of the HLBT entropy $\hat{S}$ given in Eq. (\ref{shat}) by limiting the sum to $N_{\rm t}$ and substituting to $(\mathbf{r}_i-\mathbf{r}_{n,i})^2$ and $(\mathbf{v}_i-\mathbf{v}_{n,i})^2$ in $D_i$, the quantities $\bar{d}_i^2$ and $\sigma_i^2$, respectively, defined as the square of the average interparticle distance and velocity dispersion of the underlying model at the position of the $i-$th tracer particle. The so-adapted dynamical entropy $S^*$ (minus its initial value $S^*_0$) is plotted as a function of time in Fig.~\ref{emit} for the systems discussed above (top panels) along with the corresponding tracer emittance (bottom panels).\\
\indent In all cases, the modified HLBT entropy rapidly grows on a scale of about 4 time units, reaching a local maximum that is systematically larger for smaller $N$ in the cases of collapsing models, and vice versa is smaller for the cusp steepening experiment. We observe that, for low $N$ systems (indicated by darker shades of blue, green and red), $S^*$ continues to grow even at later times after a damped oscillatory phase, with a multi power-law behaviour. By contrast, for increasingly larger values of $N$, the curve relaxes towards a nearly invariant value that is of the same order across all types of numerical experiments. This is consistent with the picture that in models with larger discreteness effects, particles initially laying on the same energy surface can explore a wider volume in phase-space, as the collective potential evolves towards a quasi stationary state.\\
\indent The emittance $\varepsilon$ has a qualitatively similar trend to that of $S^*$, however, the initial violent phases, say below $t\approx 4$ is indistinguishable for different values of $N$. We fitted the curve (see the black dashed lines in figure) in this range with the generalization of Eq. (\ref{emitfit}) 
\begin{equation}\label{emitfit2}
\varepsilon(t)=C\left({\frac{t}{N_{\rm t}}}\right)^a\exp(t\tilde{\lambda}).
\end{equation}
The fit parameters are summarized in Tab.~(\ref{tab:emitfit}). Curiously, in the collapse simulations the exponent $a$ of the power-law envelope is close to $1/2$, as one would expect from Eq. (\ref{emitfit}) recovered for example by \cite{2004PhRvS...7a4202K} for tracer particles propagating in frozen $N-$body potentials, while in the cusp growth case, the associated trend is almost linear ($a\approx 1.01$). We find that the best fit values of the inverse time scale $\tilde{\lambda}$ are of the same order of magnitude of the average Lyapunov exponent evaluated for the continuum model (i.e. without noise and friction) over the time window where the first peak of $\varepsilon$ happens. We performed additional runs with analogous initial conditions and different values of the tracer number $N_{\rm t}$ spanning from $3\times10^2$ to $2\times10^6$, finding that both $a$ and $\tilde{\lambda}$ are almost insensible to the number of tracers, provided that the latter is larger than $\approx3\times 10^3$, while the final value attained by the emittance is systematically slightly lower values of $N_{\rm t}\lesssim 3\times 10^3$ when $N\lesssim10^8$ and it is almost independent on $N_{\rm t}$ for larger sizes, as shown in the lower panel of Fig. \ref{emit_11} for a collapse with cusp growth with $N=10^{11}$. The HLBT entropy $S^*$ at fixed $N$, (as well as in the continuum limit) is even less sensible to $N_{\rm t}$, as shown in the top panel of Fig. \ref{emit_11}, at variance with what observed by \cite{2019ApJ...872...20B,2019ApJ...870..128B} for ensembles of tracers propagated in static integrable or non integrable potentials.\\ 
\begin{table}[h]
    \centering
    \caption{Fit parameters for Eq. (\ref{emitfit2})}
    \begin{tabular}{l|ccc}
        Simulation & $a$ & $\tilde{\lambda}$ & C \\
        \hline
        Self-similar collapse & 0.501 &  0.580 & 17 \\ 
        Cusp growth & 1.013  & 0.021 & 5770 \\ 
        Collapse with cusp growth & 0.498 &  0.624 & 4.736\\ 
    \end{tabular}
    \label{tab:emitfit}
\end{table}
\indent Comparing the evolution of $S^*$ and $\varepsilon$ reveals that despite both qualitatively behaving as entropies (i.e. both are increasing functions of time), the latter seems to imply an almost negligible role of the discreteness effects during the initial VR phase, while $S^*$ has systematically a markedly $N-$dependent character for the whole simulation time span. This stems from the fact that the emittance is essentially a hypersurface enclosing the available phase-space volume, while the HLBT entropy $S^*$ is a local quantity, by construction dependent on the particle density and velocity distributions, thus intrinsically more sensible to $N_{\rm t}$ scaling and discreteness effects. 
\section{Outlook and discussion}\label{remarks}
We have investigated the phase-space properties of orbit ensembles in time-dependent potentials simulating a process of VR including the $N-$dependent discreteness effects accounting for the granularity of gravitational systems. Using the stochastic integrator for independent trajectories developed by \cite{Pasquato_2020} and \cite{Di_Cintio_2019} we investigated a self-similar collapse of a cuspy system, a cusp growth in-fall and a collapse with cusp growth. Aiming at studying the influence of micro-chaos induced by the finite-$N$ nature in the case of large-$N$ models, we analyzed the energy-dependent phase-space spreading, the associated energy spectra, the distribution of orbits' largest Lyapunov exponents and their mean values as function of the system size $N$. In addition, we also studied the time evolution of two formulations of the dynamical entropy, as defined by \cite{1986MNRAS.219..285T} and \cite{1991JAP....70.1919R}.\\
\indent We find that, for sufficiently accurate samplings of the initial phase-space, the distribution of tracers moving under the effect of a time-dependent relaxing potential and discreteness effects,  here parametrized by friction and noise coefficients in the context of a Langevin stochastic model, asymptotically reproduce the energy and density distribution of the associated nearly invariant equilibrium state. Phase-space sections have a remarkably good agreement with those of full $N-$body simulations of dissipationless collapses with analogous initial conditions (see Fig.~
\ref{phase}). We speculate that, by tuning the strength of the friction coefficient $\eta$ and the fluctuation distribution $\mathcal{F}$ in Eq. (\ref{langeq}), one might improve the adiabatic squeezing scheme of \cite{2001ApJ...549..862H} to produce flattened and triaxial models starting from a spherical seed system with ergodic distribution, as in presence of discreteness effects the phase-space transport would be more efficient.\\
\indent As expected, the values of the Lyapunov exponents for fixed initial orbital energy $E_0$ are systematically lower for increasing $N$ at fixed simulation set-up, in agreement with the results by \cite{Di_Cintio_2019_revisited,Di_Cintio_2019} for direct $N$-body integrations of equilibrium models. As our method allows to probe, at least in a statistical sense, large values of $N$ in the astrophysically relevant range, we observed that there exist a system size $N_*\approx 10^{9}$ for which the trends with $N$ of some specific moments of the largest Lyapunov exponents distributions have a significant change of slope, seemingly independent on the number of tracer orbits $N_{\rm t}$ and the specific type of numerical experiment. We interpret this fact as the existence of a regime of system sizes below which the chaos induced by discreteness effects dominates the one associated with collective oscillations during the phase-mixing process after the initial phase of the collapse, and possibly at the origin of what is commonly referred to as ``incomplete violent relaxation" in numerical experiments (see \citealt{2004astro.ph..6236T,2005A&A...429..161T,2005A&A...433...57T}). This is also in agreement with the evolution of the dynamical entropies (see Fig.~\ref{emit_N}). In particular the tracer ensemble emittance $\varepsilon$ has the same behavior during the first violent phase, for all $N$ as well as for all $N_{\rm t}$ at fixed $N$, while it shows a rather clear $N$-dependence during the virial oscillation phases at later times with a distinct drift for the lower $N$ cases. This is in apparent contradiction with the argument of \cite{2017ApJ...846..125B} that a description in terms of the CBE of the VR is erroneous as it would imply a constant dynamical entropy even during violent relaxation. We conjecture that the timescale of VR might be in some cases shorter than the smallest Lyapunov timescale (i.e. the largest Lyapunov exponent) of the full $N$-body problem, but at the same time longer than the Lyapunov timescales of individual particle orbits. This implies that a given model could be relatively insensible to the discreteness effects during the fast VR phase, even though the distribution of orbits' Lyapunov exponents has a larger variance and mean for decreasing $N$ (see Figs.~\ref{lyap_N} and \ref{emit_N}). 
The origin of this and other discrepancies are essentially to be attributed to the different indicators employed to measure chaos and relaxation. Local indicators (i.e. individual orbit Lyapunov exponents and their distributions) do keep track of micro-chaos, while integrated quantities like the HLBT entropy are more sensitive to collective processes. \\
\indent More broadly, this point likely connects to the argument by \cite{2005MNRAS.361..385A} that violent relaxation is largely insensitive to the fine-grained evolution of $f$ in phase space, even within the collisionless CBE picture. As a result, any instance of violent relaxation is inherently incomplete. The subsequent, more gradual fluctuations that steer the system toward a quasi-equilibrium state must therefore be understood only in a coarse-grained sense, involving a discretized phase-space density $\bar{f}$, as in \cite{1998MNRAS.300..981C} and \cite{2022MNRAS.512.3015B}.
\begin{acknowledgements}
We thank the anonymous referee for insightful suggestions that improved the manuscript. P.F.D.C. wishes to acknowledge funding by ``Fondazione Cassa di Risparmio di Firenze'' under the project {\it HIPERCRHEL} for the use of high performance computing resources at the University of Firenze. A.A.T. acknowledges support from the Horizon Europe research and innovation programs under the Marie Sk\l{}odowska-Curie grant agreement no. 101103134.
\end{acknowledgements}
\bibliographystyle{aa} 
\bibliography{bibthesis} 

\begin{thebibliography}{88}
\expandafter\ifx\csname natexlab\endcsname\relax\def\natexlab#1{#1}\fi

\bibitem[{{Arad} \& {Lynden-Bell}(2005)}]{2005MNRAS.361..385A}
{Arad}, I. \& {Lynden-Bell}, D. 2005, \mnras, 361, 385

\bibitem[{Arnold {et~al.}(1986)Arnold, Kliemann, \&
  Oeljeklaus}]{10.1007/BFb0076835}
Arnold, L., Kliemann, W., \& Oeljeklaus, E. 1986, in Lyapunov Exponents, ed.
  L.~Arnold \& V.~Wihstutz (Berlin, Heidelberg: Springer Berlin Heidelberg),
  85--125

\bibitem[{{Baes} \& {Dejonghe}(2021)}]{2021A&A...653A.140B}
{Baes}, M. \& {Dejonghe}, H. 2021, \aap, 653, A140

\bibitem[{{Banik} {et~al.}(2022){Banik}, {Weinberg}, \& {van den
  Bosch}}]{2022ApJ...935..135B}
{Banik}, U., {Weinberg}, M.~D., \& {van den Bosch}, F.~C. 2022, \apj, 935, 135

\bibitem[{{Barbieri} {et~al.}(2022){Barbieri}, {Di Cintio}, {Giachetti},
  {Simon-Petit}, \& {Casetti}}]{2022MNRAS.512.3015B}
{Barbieri}, L., {Di Cintio}, P., {Giachetti}, G., {Simon-Petit}, A., \&
  {Casetti}, L. 2022, \mnras, 512, 3015

\bibitem[{{Benettin} {et~al.}(1976){Benettin}, {Galgani}, \&
  {Strelcyn}}]{1976PhRvA..14.2338B}
{Benettin}, G., {Galgani}, L., \& {Strelcyn}, J.-M. 1976, \pra, 14, 2338

\bibitem[{{Beraldo e Silva} {et~al.}(2017){Beraldo e Silva}, {de Siqueira
  Pedra}, {Sodr{\'e}}, {Perico}, \& {Lima}}]{2017ApJ...846..125B}
{Beraldo e Silva}, L., {de Siqueira Pedra}, W., {Sodr{\'e}}, L., {Perico}, E.
  L.~D., \& {Lima}, M. 2017, \apj, 846, 125

\bibitem[{{Beraldo e Silva} {et~al.}(2019{\natexlab{a}}){Beraldo e Silva}, {de
  Siqueira Pedra}, \& {Valluri}}]{2019ApJ...872...20B}
{Beraldo e Silva}, L., {de Siqueira Pedra}, W., \& {Valluri}, M.
  2019{\natexlab{a}}, \apj, 872, 20

\bibitem[{{Beraldo e Silva} {et~al.}(2019{\natexlab{b}}){Beraldo e Silva}, {de
  Siqueira Pedra}, {Valluri}, {Sodr{\'e}}, \& {Bru}}]{2019ApJ...870..128B}
{Beraldo e Silva}, L., {de Siqueira Pedra}, W., {Valluri}, M., {Sodr{\'e}}, L.,
  \& {Bru}, J.-B. 2019{\natexlab{b}}, \apj, 870, 128

\bibitem[{{Beraldo e Silva} {et~al.}(2024){Beraldo e Silva}, {Valluri},
  {Vasiliev}, {Hattori}, {de Siqueira Pedra}, \&
  {Daniel}}]{2024arXiv240707947B}
{Beraldo e Silva}, L., {Valluri}, M., {Vasiliev}, E., {et~al.} 2024, arXiv
  e-prints, arXiv:2407.07947

\bibitem[{{Binney} \& {Tremaine}(2008)}]{binney}
{Binney}, J. \& {Tremaine}, S. 2008, {Galactic Dynamics: Second Edition}

\bibitem[{{Breden} {et~al.}(2024){Breden}, {Chu}, {Lamb}, \&
  {Rasmussen}}]{2024arXiv241107064B}
{Breden}, M., {Chu}, H., {Lamb}, J. S.~W., \& {Rasmussen}, M. 2024, arXiv
  e-prints, arXiv:2411.07064

\bibitem[{{Campa} {et~al.}(2014){Campa}, {Dauxois}, {Fanelli}, \&
  {Ruffo}}]{2014plis.book.....C}
{Campa}, A., {Dauxois}, T., {Fanelli}, D., \& {Ruffo}, S. 2014, {Physics of
  long-range interacting systems}

\bibitem[{{Casetti} {et~al.}(2000){Casetti}, {Pettini}, \&
  {Cohen}}]{2000PhR...337..237C}
{Casetti}, L., {Pettini}, M., \& {Cohen}, E.~G.~D. 2000, \physrep, 337, 237

\bibitem[{{Cerruti-Sola} \& {Pettini}(1995)}]{1995PhRvE..51...53C}
{Cerruti-Sola}, M. \& {Pettini}, M. 1995, \pre, 51, 53

\bibitem[{{Chandrasekhar}(1943)}]{chandra}
{Chandrasekhar}, S. 1943, ApJ, 97, 255

\bibitem[{{Chandrasekhar} \& {von Neumann}(1943)}]{chandra_vonnumann}
{Chandrasekhar}, S. \& {von Neumann}, J. 1943, ApJ, 97, 1

\bibitem[{{Chavanis}(1998)}]{1998MNRAS.300..981C}
{Chavanis}, P.-H. 1998, \mnras, 300, 981

\bibitem[{{Ciotti}(2010)}]{2010AIPC.1242..117C}
{Ciotti}, L. 2010, in American Institute of Physics Conference Series, Vol.
  1242, Plasmas in the Laboratory and the Universe: Interactions, Patterns, and
  Turbulence, ed. G.~{Bertin}, F.~{de Luca}, G.~{Lodato}, R.~{Pozzoli}, \&
  M.~{Rom{\'e}} (AIP), 117--128

\bibitem[{{Dehnen}(1993)}]{1993MNRAS.265..250D}
{Dehnen}, W. 1993, \mnras, 265, 250

\bibitem[{{Dehnen}(2005)}]{2005MNRAS.360..892D}
{Dehnen}, W. 2005, \mnras, 360, 892

\bibitem[{{Di Cintio}(2024)}]{2024A&A...683A.254D}
{Di Cintio}, P. 2024, \aap, 683, A254

\bibitem[{Di~Cintio \& Casetti(2019)}]{Di_Cintio_2019_revisited}
Di~Cintio, P. \& Casetti, L. 2019, Monthly Notices of the Royal Astronomical
  Society, 489, 5876–5888

\bibitem[{{Di Cintio} \& {Casetti}(2020)}]{2020MNRAS.494.1027D}
{Di Cintio}, P. \& {Casetti}, L. 2020, \mnras, 494, 1027

\bibitem[{Di~Cintio \& Casetti(2020)}]{Di_Cintio_casetti}
Di~Cintio, P. \& Casetti, L. 2020, Proceedings of the International
  Astronomical Union, 14, 426–429

\bibitem[{{Di Cintio} {et~al.}(2013){Di Cintio}, {Ciotti}, \&
  {Nipoti}}]{2013MNRAS.431.3177D}
{Di Cintio}, P., {Ciotti}, L., \& {Nipoti}, C. 2013, \mnras, 431, 3177

\bibitem[{Di~Cintio {et~al.}(2020)Di~Cintio, Ciotti, \&
  Nipoti}]{Di_Cintio_2019}
Di~Cintio, P., Ciotti, L., \& Nipoti, C. 2020, Proceedings of the International
  Astronomical Union, 14, 93–96

\bibitem[{{Di Cintio} \& {Trani}(2025)}]{2025A&A...693A..53D}
{Di Cintio}, P. \& {Trani}, A.~A. 2025, \aap, 693, A53

\bibitem[{{El-Zant}(2002)}]{2002MNRAS.331...23E}
{El-Zant}, A.~A. 2002, \mnras, 331, 23

\bibitem[{{El-Zant} {et~al.}(2019){El-Zant}, {Everitt}, \&
  {Kassem}}]{2019MNRAS.484.1456E}
{El-Zant}, A.~A., {Everitt}, M.~J., \& {Kassem}, S.~M. 2019, \mnras, 484, 1456

\bibitem[{{Goodman} {et~al.}(1993){Goodman}, {Heggie}, \&
  {Hut}}]{1993ApJ...415..715G}
{Goodman}, J., {Heggie}, D.~C., \& {Hut}, P. 1993, \apj, 415, 715

\bibitem[{{Gurzadian} \& {Savvidy}(1986)}]{GurzadianSavvidy}
{Gurzadian}, V.~G. \& {Savvidy}, G.~K. 1986, A\&A, 160, 203

\bibitem[{{Gurzadyan} \& {Kocharyan}(2009)}]{2009A&A...505..625G}
{Gurzadyan}, V.~G. \& {Kocharyan}, A.~A. 2009, \aap, 505, 625

\bibitem[{{He}(2013)}]{2013IJMPB..2761011H}
{He}, P. 2013, International Journal of Modern Physics B, 27, 1361011

\bibitem[{{Hemsendorf} \& {Merritt}(2002)}]{2002ApJ...580..606H}
{Hemsendorf}, M. \& {Merritt}, D. 2002, \apj, 580, 606

\bibitem[{{Holley-Bockelmann} {et~al.}(2001){Holley-Bockelmann}, {Mihos},
  {Sigurdsson}, \& {Hernquist}}]{2001ApJ...549..862H}
{Holley-Bockelmann}, K., {Mihos}, J.~C., {Sigurdsson}, S., \& {Hernquist}, L.
  2001, \apj, 549, 862

\bibitem[{{Holtsmark}(1919)}]{1919AnP...363..577H}
{Holtsmark}, J. 1919, Annalen der Physik, 363, 577

\bibitem[{{Kandrup}(1987)}]{1987MNRAS.225..995K}
{Kandrup}, H.~E. 1987, \mnras, 225, 995

\bibitem[{{Kandrup}(1996)}]{1996kandrup}
{Kandrup}, H.~E. 1996, in Proceedings of the Seventh Marcel Grossman Meeting on
  recent developments in theoretical and experimental general relativity,
  gravitation, and relativistic field theories, ed. R.~T. {Jantzen}, G.~{Mac
  Keiser}, \& R.~{Ruffini}, Vol.~7, 167--182

\bibitem[{{Kandrup}(1998{\natexlab{a}})}]{1998NYASA.848...28K}
{Kandrup}, H.~E. 1998{\natexlab{a}}, Annals of the New York Academy of
  Sciences, 848, 28

\bibitem[{{Kandrup}(1998{\natexlab{b}})}]{1998ApJ...500..120K}
{Kandrup}, H.~E. 1998{\natexlab{b}}, \apj, 500, 120

\bibitem[{{Kandrup} {et~al.}(2005){Kandrup}, {Bohn}, {Kishek}, {O'Shea},
  {Reiser}, \& {Sideris}}]{2005NYASA1045...12K}
{Kandrup}, H.~E., {Bohn}, C.~L., {Kishek}, R.~A., {et~al.} 2005, Annals of the
  New York Academy of Sciences, 1045, 12

\bibitem[{{Kandrup} \& {Novotny}(2004)}]{2004CeMDA..88....1K}
{Kandrup}, H.~E. \& {Novotny}, S.~J. 2004, Celestial Mechanics and Dynamical
  Astronomy, 88, 1

\bibitem[{{Kandrup} \& {Sideris}(2001)}]{2001PhRvE..64e6209K}
{Kandrup}, H.~E. \& {Sideris}, I.~V. 2001, \pre, 64, 056209

\bibitem[{Kandrup \& Sideris(2003)}]{Kandrup_2003}
Kandrup, H.~E. \& Sideris, I.~V. 2003, The Astrophysical Journal, 585,
  244–249

\bibitem[{{Kandrup} {et~al.}(2004){Kandrup}, {Sideris}, \&
  {Bohn}}]{2004PhRvS...7a4202K}
{Kandrup}, H.~E., {Sideris}, I.~V., \& {Bohn}, C.~L. 2004, Physical Review
  Accelerators and Beams, 7, 014202

\bibitem[{{Kandrup} {et~al.}(2003){Kandrup}, {Vass}, \&
  {Sideris}}]{kandrupvasssideris}
{Kandrup}, H.~E., {Vass}, I.~M., \& {Sideris}, I.~V. 2003, MNRAS, 341, 927

\bibitem[{{Kang} \& {He}(2011)}]{2011A&A...526A.147K}
{Kang}, D.~B. \& {He}, P. 2011, \aap, 526, A147

\bibitem[{Kantz \& Schreiber(2004)}]{kantz2004nonlinear}
Kantz, H. \& Schreiber, T. 2004, Nonlinear Time Series Analysis, Cambridge
  nonlinear science series (Cambridge University Press)

\bibitem[{Kim \& Littlejohn(1995)}]{505881}
Kim, K.-J. \& Littlejohn, R. 1995, in Proceedings Particle Accelerator
  Conference, Vol.~5, 3358--3360 vol.5

\bibitem[{{Kolmogorov}(1958)}]{kolmogorov0}
{Kolmogorov}, A. 1958, Doklady Akademii Nauk SSSR, 119, 861

\bibitem[{{Kolmogorov}(1959)}]{kolmogorov}
{Kolmogorov}, A. 1959, Doklady Akademii Nauk SSSR, 124, 754

\bibitem[{Laffargue {et~al.}(2016)Laffargue, Tailleur, \& van
  Wijland}]{Laffargue_2016}
Laffargue, T., Tailleur, J., \& van Wijland, F. 2016, Journal of Statistical
  Mechanics: Theory and Experiment, 2016, 034001

\bibitem[{{Leonenko} {et~al.}(2008){Leonenko}, {Pronzato}, \&
  {Savani}}]{2008arXiv0810.5302L}
{Leonenko}, N., {Pronzato}, L., \& {Savani}, V. 2008, Annals of Statistics, 36,
  2153

\bibitem[{{Levin} {et~al.}(2014){Levin}, {Pakter}, {Rizzato}, {Teles}, \&
  {Benetti}}]{2014PhR...535....1L}
{Levin}, Y., {Pakter}, R., {Rizzato}, F.~B., {Teles}, T.~N., \& {Benetti}, F.
  P.~C. 2014, \physrep, 535, 1

\bibitem[{Lichtenberg \& Lieberman(2013)}]{lichtenberg2013regular}
Lichtenberg, A. \& Lieberman, M. 2013, Regular and Chaotic Dynamics, Applied
  Mathematical Sciences (Springer New York)

\bibitem[{{Londrillo} {et~al.}(1991){Londrillo}, {Messina}, \&
  {Stiavelli}}]{1991MNRAS.250...54L}
{Londrillo}, P., {Messina}, A., \& {Stiavelli}, M. 1991, \mnras, 250, 54

\bibitem[{{Lorenz}(1963)}]{DeterministicNonperiodicFlow}
{Lorenz}, E.~N. 1963, Journal of Atmospheric Sciences, 20, 130

\bibitem[{Lynden-Bell(1967)}]{Lynden-Bell:1966zjv}
Lynden-Bell, D. 1967, Mon. Not. Roy. Astron. Soc., 136, 101

\bibitem[{{Madsen}(1987)}]{1987ApJ...316..497M}
{Madsen}, J. 1987, \apj, 316, 497

\bibitem[{Mannella(2004)}]{Mannella_2004}
Mannella, R. 2004, Physical Review E, 69

\bibitem[{{Merritt}(2005)}]{2005NYASA1045....3M}
{Merritt}, D. 2005, Annals of the New York Academy of Sciences, 1045, 3

\bibitem[{{Miller}(1964)}]{1964ApJ...140..250M}
{Miller}, R.~H. 1964, \apj, 140, 250

\bibitem[{{Miller}(1971)}]{1971JCoPh...8..449M}
{Miller}, R.~H. 1971, Journal of Computational Physics, 8, 449

\bibitem[{{Nakamura}(2000)}]{2000ApJ...531..739N}
{Nakamura}, T.~K. 2000, \apj, 531, 739

\bibitem[{{Nipoti} {et~al.}(2007){Nipoti}, {Londrillo}, \&
  {Ciotti}}]{2007ApJ...660..256N}
{Nipoti}, C., {Londrillo}, P., \& {Ciotti}, L. 2007, \apj, 660, 256

\bibitem[{{Padmanabhan}(1990)}]{1990PhR...188..285P}
{Padmanabhan}, T. 1990, \physrep, 188, 285

\bibitem[{Pasquato \& Di~Cintio(2020)}]{Pasquato_2020}
Pasquato, M. \& Di~Cintio, P. 2020, Astronomy \& Astrophysics, 640, A79

\bibitem[{{Pesin}(1977)}]{1977RuMaS..32...55P}
{Pesin}, Y.~B. 1977, Russian Mathematical Surveys, 32, 55

\bibitem[{{Reiser}(1991)}]{1991JAP....70.1919R}
{Reiser}, M. 1991, Journal of Applied Physics, 70, 1919

\bibitem[{{Severne} \& {Luwel}(1980)}]{1980Ap&SS..72..293S}
{Severne}, G. \& {Luwel}, M. 1980, \apss, 72, 293

\bibitem[{Sideris \& Kandrup(2002)}]{Sideris_2002}
Sideris, I.~V. \& Kandrup, H.~E. 2002, Physical Review E, 65

\bibitem[{Sideris \& Kandrup(2004)}]{Sideris_2004}
Sideris, I.~V. \& Kandrup, H.~E. 2004, The Astrophysical Journal, 602,
  678–684

\bibitem[{{Sinai}(1959)}]{sinai}
{Sinai}, Y. 1959, Doklady Akademii Nauk SSSR, 124, 768

\bibitem[{Skokos(2009)}]{Skokos_2009}
Skokos, C. 2009, The Lyapunov Characteristic Exponents and Their Computation
  (Springer Berlin Heidelberg), 63–135

\bibitem[{{Soker}(1996)}]{1996ApJ...457..287S}
{Soker}, N. 1996, \apj, 457, 287

\bibitem[{{Struckmeier}(1996)}]{1996PhRvE..54..830S}
{Struckmeier}, J. 1996, \pre, 54, 830

\bibitem[{{Sylos Labini}(2012)}]{2012MNRAS.423.1610S}
{Sylos Labini}, F. 2012, \mnras, 423, 1610

\bibitem[{{Tremaine} {et~al.}(1986){Tremaine}, {Henon}, \&
  {Lynden-Bell}}]{1986MNRAS.219..285T}
{Tremaine}, S., {Henon}, M., \& {Lynden-Bell}, D. 1986, \mnras, 219, 285

\bibitem[{{Tremaine} {et~al.}(1994){Tremaine}, {Richstone}, {Byun}, {Dressler},
  {Faber}, {Grillmair}, {Kormendy}, \& {Lauer}}]{1994AJ....107..634T}
{Tremaine}, S., {Richstone}, D.~O., {Byun}, Y.-I., {et~al.} 1994, \aj, 107, 634

\bibitem[{{Trenti} \& {Bertin}(2004)}]{2004astro.ph..6236T}
{Trenti}, M. \& {Bertin}, G. 2004, arXiv e-prints, astro

\bibitem[{{Trenti} \& {Bertin}(2005)}]{2005A&A...429..161T}
{Trenti}, M. \& {Bertin}, G. 2005, \aap, 429, 161

\bibitem[{{Trenti} {et~al.}(2005){Trenti}, {Bertin}, \& {van
  Albada}}]{2005A&A...433...57T}
{Trenti}, M., {Bertin}, G., \& {van Albada}, T.~S. 2005, \aap, 433, 57

\bibitem[{{van Albada}(1982)}]{1982MNRAS.201..939V}
{van Albada}, T.~S. 1982, \mnras, 201, 939

\bibitem[{{Vass} {et~al.}(2003){Vass}, {Kandrup}, \&
  {Terzi{\'c}}}]{kandrupvass}
{Vass}, I.~M., {Kandrup}, H.~E., \& {Terzi{\'c}}, B. 2003, in American
  Astronomical Society Meeting Abstracts, Vol. 203, American Astronomical
  Society Meeting Abstracts, 91.11

\bibitem[{{Vesperini}(1992)}]{1992A&A...266..215V}
{Vesperini}, E. 1992, \aap, 266, 215

\bibitem[{{Wolf} {et~al.}(1985){Wolf}, {Swift}, {Swinney}, \&
  {Vastano}}]{1985PhyD...16..285W}
{Wolf}, A., {Swift}, J.~B., {Swinney}, H.~L., \& {Vastano}, J.~A. 1985, Physica
  D Nonlinear Phenomena, 16, 285

\bibitem[{{Worrakitpoonpon}(2024)}]{2024PhRvE.109e4118W}
{Worrakitpoonpon}, T. 2024, \pre, 109, 054118

\end{thebibliography}
\end{document}